\documentstyle[preprint,aps]{revtex}
\input epsf.tex
\def\DESepsf(#1 width #2){\epsfxsize=#2 \epsfbox{#1}}
\begin{document}
\preprint{\vbox{\hbox{}}}
\draft
\title{The Neutrino Magnetic Moment Induced by Leptoquarks}
\author{Chun-Khiang Chua and W-Y. P. Hwang}
\address{
Department of Physics, National Taiwan University,\\
Taipei, Taiwan, Republic of China\\
}
\date{November, 1998}
\maketitle
\begin{abstract}
Allowing leptoquarks to interact with both right-handed and left-handed
neutrinos (i.e., ``non-chiral'' leptoquarks), we show that a non-zero
neutrino magnetic moment can arise naturally. Although the mass of the
non-chiral vector leptoquark that couples to the first generation fermions is
constrained severely by universality of the $\pi^+$ leptonic decays and is
found to be greater than 50 TeV, the masses of the second and third generation
non-chiral vector leptoquarks may evade such constraint and may in general be
in the range of $1\sim 100$ TeV. With reasonable input mass and coupling
values, we find that the neutrino magnetic moment due to the second generation
leptoquarks is of the order of $10^{-12}\sim 10^{-16}\,\mu_{\rm B}$ while that
caused by the third generation leptoquarks, being enhanced significantly by the
large top quark mass, is in the range of $10^{-10}\sim 10^{-14}\,\mu_{\rm B}$.
\end{abstract}


\preprint{\vbox{\hbox{}}}

\preprint{\vbox{\hbox{}}}


\section{Introduction}

Existence of a non-zero neutrino magnetic moment has long been a concern of
great interest, since it can have an observable laboratory effect such as
neutrino-charge lepton elastic scattering, $e^+e^-\rightarrow
\nu{\bar\nu}\gamma$, and also some important astrophysics effect, 
such as cooling of SN 1987A, cooling of helium stars, etc. 
It is likely that neutrinos may have a small but nonvanishing mass;
for various bounds on magnetic moments and masses, see \cite{pdg} and
\cite{superK}. 
Within the framework of the standard model, a nonzero neutrino
mass usually imply a nonzero magnetic moment. It has been shown 
that \cite{fuji}, for a massive neutrino,
\begin{equation}                                           
\mu^{SM}_\nu={3 e G_F m_\nu\over 8\pi^2\sqrt 2}            
            =3.2\times10^{-19}\, m_\nu({\rm eV})\, \mu_B,                      
\end{equation} 
where $\mu_B$ is the Bohr magneton.

In models beyond the standard model, right-handed neutrinos are often included
in interactions (for a review e.g. \cite{moha}), so that we need not depend on
a nonzero neutrino mass to generate a nonzero magnetic moment. In this article,
we consider the possibility of using
leptoquark interactions to generate a nonzero neutrino magnetic moment.
In many unification models, such as SU(5), S0(10), etc., 
one often put quarks and leptons
into the same multiplet, so that leptoquarks arise naturally for connecting
different components within the same multiplet. What makes a leptoquark unique
and interesting is that it couples simultaneously to both a lepton and a
quark. This may help generating a nonzero neutrino magnetic interaction.  
Specifically, when a top quark involves in the loop diagram, 
its mass provide a large enhancement for the neutrino magnetic moment.
(In such diagram, a massless neutrino needs some massive internal fermion to
flip its chirality, giving rise to some nonzero magnetic moment.) 

We add right-handed neutrinos in the general renormalizable lagrangian of 
leptoquarks. Owing to existence of lepton numbers which recognize the
generation, we distinguish leptoquarks by its generation quantum number, but
this may induce four-fermion interactions which may enhance some helicity
suppressed process such as $\pi^+\rightarrow {e^+} \nu$ to the extent that 
leptonic universality may even be violated. This usually gives a tight 
constraint on the leptoquark \cite{constraint}. For non-chiral vector
leptoquarks of electromagnetic strength coupling, this corresponds 
to having a mass heavier than 50 TeV for the first generation
leptoquark. With such a heavy leptoquark, we still find a nonzero neutrino
magnetic moment $\mu_{\nu}$ up to $10^{-18}\,\mu_{B}$. For the second and third
generation leptoquarks, their masses are not severely constrained by the above
process. Assuming their mass lying somewhere between  $1\sim 100$ TeV, 
we obtain $\mu_{\nu}$ of order $10^{-12}\sim 10^{-16}\,\mu_{\rm B}$ (from the
second generation leptoquark) and $10^{-10}\sim 10^{-14}\,\mu_{\rm B}$ (from
the third generation leptoquark), respectively. Such predictions may already 
have some observable effects such as those mentioned earlier.

\section{Neutrino magnetic moment in models with leptoquarks}

Leptoquarks arise naturally in many unification models which attempt to put
quarks and leptons in the same multiplet. There are scalar and vector
leptoquarks which may couple to left- and right-handed 
neutrinos at the same time, but only vector leptoquarks can couple to the
upper component of the quark SU(2) doublet. The heaviness of the
top quark may enhance the neutrino magnetic moment once we use the vector
leptoquark to connect to the quark doublet. Of course, there are  
subtleties regarding renormalization of the vector leptoquark which may be
treated in a way similar to gauge bosons. In our calculation, we adopt 
Feynman rules in the $R_{\xi}$-gauge and take $\xi \to\infty$ at the end 
of the calculation while neglecting all unphysical particles in the
$R_\xi$-gauge. (This is a step which has often been employed in nonabelian
gauge theories.)

We begin our analysis by constructing a general renormalizable
lagrangian for the quark-lepton-leptoquark coupling. Following \cite{buch},
we demand such action to be an SU(3)$\times$SU(2)$\times$U(1) invariant
which conserves the baryon and lepton numbers but, in addition to \cite{buch} 
we add terms which couple to right-handed neutrinos. For leptoquarks with the
fermion number $F\equiv 3B+L=0$,

\begin{eqnarray}
{\cal L}_{F=0}&=& (g_{1L} {\bar Q_L}\gamma^\mu L_L+
                 g_{1R} {\bar D_R}\gamma^\mu l_R+
                 g^u_{1\nu} {\bar U_R}\gamma^\mu \nu_R)
                               \, V^{({2\over3})}_{1\mu}\nonumber\\
     &&+(g^d_{2L} {\bar D_R} L^i_L i\tau_{2\,ij}+g_{2\nu} 
                          {\bar Q_{L\,j}} \nu_R)\, S^{j({1\over6})}_2\nonumber\\
     &&+(g^u_{2L} {\bar U_R} L^i_L i\tau_{2\,ij}+g_{2R}
                          {\bar Q_{L\,j}} l_R)\, S^{j({7\over6})}_2\nonumber\\
     &&+g_{3L} {\bar Q_L} {\vec \tau} \gamma^\mu L_L\, {\vec V}^{({2\over3})}
          _{3\mu}
     +g^u_{1R} {\bar U_R} \gamma^\mu l_R\, V^{({5\over3})}  
          _{1\mu}
 +g^d_{1\nu} {\bar D_R} \gamma^\mu \nu_R\, V^{(-{1\over3})}    
          _{1\mu}\nonumber\\
     &&+{\rm c. c.},
\end{eqnarray}
and, for $F=\pm2$,
\begin{eqnarray}
{\cal L}_{F=2}&=& (h_{2L} {\bar u_R}^c\gamma^\mu L^i_Li\tau_{2ij}
                + h_{2\nu} {\bar Q_L}^ci\tau_{2ij}\gamma^\mu\nu_R)
                               \, V^{(-{1\over6})}_{2\mu}\nonumber\\
&&+(h_{1L} {\bar Q_L}^{i\,c} i\tau_{2\,ij} L^j_L
      +h_{1R} {\bar U_R}^c l_R
      +h_{1\nu} {\bar D_R}^c \nu_R)\, S^{({1\over3})}_1\nonumber\\
&&+(h_{2L} {\bar D_R}^c\gamma^\mu L^i_L i\tau_{2\,ij}
        +h_{2R}{\bar Q_L}^{i\,c}i\tau_{2ij}\gamma^\mu l_R)\,
                                           V^{j({5\over6})}_{2\mu}\nonumber\\
&&+h_{3L} {\bar Q_L}^{c\,i}{\vec\tau}i\tau_{2ij} L_L\, S^{({1\over3})}_{3}
   +h_{1R} {\bar D_R}^c l_R\, S^{(-{4\over3})}_1
                  +h_{1\nu} {\bar U_R}^c \nu_R\, S^{({2\over3})}_1 \nonumber\\
&&+{\rm c. c.}.
\end{eqnarray}
The notation adopted above is self explanatory; for example, 
$S,\,V$ denotes scalar and vector leptoquarks respectively,
the superscript is its average electric charge or the hypercharge $Y$,
and the subscript of a leptoquark denotes which      
SU(2) multiplet it is in, and the generation index is suppressed.    
From here it is clear that
among those leptoquarks that couple to neutrinos of both chiralities,
a radiative $\nu\nu\gamma$ diagram with the exchange of 
a virtual $U$-type quark can proceed only when accompanied by
a vector leptoquark, namely $V^{({2\over3})}_{1\mu}$ in 
${\cal L}_{F=0}$ or $V^{(-{1\over6})}_{2\mu}$ in ${\cal L}_{F=2}$; on the other
hand, the exchange of a virtual $D$-type quark can proceed only with the scalar
leptoquark, namely $S^{({1\over6})}_{2}$ in ${\cal L}_{F=0}$ or
$S^{({1\over3})}_{1}$ in ${\cal L}_{F=2}$. 
Note that we do not consider mixing between different leptoquarks due to Higgs 
interactions, which will introduce additional parameters.
The diagram in question is shown explicitly in Fig. 1.

Given these couplings, it is straightforward to calculate induced neutrino
magnetic moments via one loop diagrams. To see that the heavy top quark mass
can enhance the prediction, we calculate the $\nu\nu\gamma$ diagram with the
exchange of up-type quark and $V^{({2\over3})}_{1\mu}$ 
i.e. the first term in ${\cal L}_{F=0}$.
As one of the standard methods to treat loop diagrams involving massive vecter
particles, we use Feynman rules in the $R_\xi$-gauge and take $\xi \to \infty$
at the end of calculation while neglecting any unphysicical particle. 
In addition to minimum substitution, we add the term 
$e\,Q_v\,V^{\dag}_\mu\,V_\nu\,F^{\mu\,\nu}$
in the lagrangian,
such that the whole $VV\gamma$ coupling is in a form similar to the non-abelian
$WW\gamma$ type coupling,
and the procedure results in a finite limit under
$\xi\rightarrow\infty$. We obtain, with all couplings chosen to be real,

\begin{eqnarray}
{\cal L}^{eff} &= &-{e\over 2 m_e} {\bar \nu} 
                   ({\,\,\sigma^{\mu\nu} \over 2})\nu
                      F_{\mu\nu}\,F_2,\,\,\,\,\,
                   |\mu_\nu|={e\over 2 m_e} F_2  \\
F_2\,\,\,\,&=&{1\over16\pi^2}{2m_e\over M}\lbrace{g^u_L g_\nu\, m_q\over M_v}
        \lbrack Q_q(\,f_1(a)+f_2(a)\,)+Q_v(\,f_3(a)+f_4(a)\,)\rbrack\nonumber\\
    & &\,\,+(\,g_L^2+g^{u2}_\nu)\,{m_\nu\over M_v}\lbrack Q_q(\,g_1(a)+g_2(a)\,)
                                      +Q_v(\,g_3(a)+g_4(a)\,)\rbrack\rbrace,
\end{eqnarray}

where $a=m_q^2/M_v^2$, $Q_q=-Q_v=2/3$, and $e>0$,
while $f_i$ and $g_i$ are given by
\begin{eqnarray}
f_1(a)&=&{2(-1+a^2-2a {\rm log}(a))\over (a-1)^3},
\nonumber\\
f_2(a)&=&-{a(3-4a+a^2+2 {\rm log}(a))\over 2(a-1)^3},
\nonumber\\
f_3(a)&=&-{3(-1+4a-3a^2+2a^2 {\rm log}(a))\over 2(a-1)^3},
\nonumber\\
f_4(a)&=&-1/2,
\nonumber\\                                                                     
g_1(a)&=&{(-4-5a^3+9a+6a(2a-1) {\rm log}(a))\over 6(a-1)^4},
\nonumber\\
g_2(a)&=&{a(3-4a+a^2+2 {\rm log}(a))\over 4(a-1)^3},
\nonumber\\
g_3(a)&=&{(7-33a+57a^2-31a^3+6a^2(3a-1) {\rm log}(a))\over 12(a-1)^4},
\nonumber\\
g_4(a)&=&{(2-6a+15a^2-14a^3+3a^4+6a^2 {\rm log}(a))\over 12(a-1)^4}.
\end{eqnarray}
Note that one obtains the desirable chiral structure for the magnetic moment
interaction by two different ways: the first is to have an odd number of mass
insertions of the quark mass term, giving rise to the first term in $F_2$;
the other way is by the neutrino mass term, resulting in the second term of
$F_2$. There are two advantages with the first scenario. First of all, one can
obtain a nonzero magnetic moment without restricted by the very light neutrino
mass. Second, one may have a prediction enhanced considerably by the heavy top 
quark mass.

\section{Constraints and Numerical Results}

Before working out numerical predictions, we need to consider the constrains 
arising from the leptonic decays of the pseudoscalar meson, such as
$\pi^+\rightarrow{e^+}\nu$\cite{constraint}. Intergrating out $V^{({2\over3})}$
and performing Fierz reordering, we obtain ${\cal L}_{eff}$ relevant
to the leptonic decay of a pseudoscalar meson,
\begin{eqnarray}
{\cal L}_{eff}&={1\over M^2_v}&(2 g^*_{1L}g_{1R}{\bar D_R}U_L{\bar \nu_L}l_R
              +2 g^{u*}_{1\nu}g_{1L}{\bar D_L}U_R{\bar \nu_R}l_L\nonumber\\
 & &-g^*_{1L}g_{1L}{\bar D_L}\gamma^\mu U_L{\bar \nu_L}\gamma_\mu l_L
 -g^{u*}_{1\nu}g_{1R}{\bar D_R}\gamma^\mu U_R{\bar \nu_R}\gamma_\mu l_R)
 +{\rm c. c.}. 
\end{eqnarray} 
We consider the universality constraint arising from the $\pi^+$ leptonic
decay, and neglect the neutrino mass contribution.
Define $R=Br(\pi^+\rightarrow e^+\nu)/Br(\pi^+\rightarrow \mu^+\nu)$.
The first and third terms of ${\cal L}_{eff}$ 
have interference with the standard model Fermi 
interaction. This is an order of $1/M^2_v$ correction to $R$,
while the other term is a correction of order $1/M^4_v$. Furthermore,
the first term which is scalar coupling is enhanced by  
a factor of $m^2_{\pi}/((m_u+m_d)m_e)$ so this is the dominent term to
constrain the mass of the leptoquark. We assume $g^{u}_{1\nu}=g_{1L}=g_{1R}=g$
which is a natural assumption for the vecter leptoquark.
We obtain
\begin{equation}
R^{exp}=R^{sm} (1+2 {m^2_{\pi}\over m_e (m_u+m_d)}
           (-{g^*_{1L}g_{1R}\over {\sqrt2} M^2_v G_F})),
\end{equation}
where experimental average $R^{exp}=(1.230\pm0.004)\times10^{-4}$\cite{pdg},
and standard model calculation $R^{sm}=(1.2352\pm0.0005)\times10^{-4}$
\cite{sirlin}. This correspond to 
\begin{equation}                                                                
M_v>g\, m_{\pi} {\sqrt{{\sqrt2}\over 0.0075 G_F m_e (m_u+m_d)}}
   \sim 50 ({g\over e}){\rm TeV}.
\end{equation}
For a coupling of the electromagnetic strength, this correspond to having the
vector leptoquark with a mass greater than 50 TeV for the first generation.
This constraint is in fact more servere than what we may obtain from the atomic
parity vialation experiment, which we shall ignore in this paper.
For the second and third generations leptoquarks, 
there is no direct restriction from the universality of the $\pi$ leptonic
decay, nor from the atomic parity violation experiment. Nevertheless,  
one can find various lower bounds for the leptoquark mass \cite{pdg},
from direct searches at the HERA $ep$ collider,
the Tevatron $p\bar p$ collider, and at the LEP $e^+e^-$ collider.
Typical bounds from direct searches is about few hundreds GeV,
while the bounds form indirect searches are given in \cite{indirect}.
We shall consider a leptoquark mass in the general range of TeV's.

For the reason of comparisons, let us recall briefly some of the upper limit
obtained from the leptonic scattering such as elastic $\nu$($\bar\nu$) 
with $l^+$($l^-$), $e^+e^-\rightarrow\nu{\bar\nu}\gamma$, etc.,
and also from the astrophysical processes such as cooling of helium stars,
red giant luminosity and so on \cite{pdg}. As a reference point, we recall the
standard model formula on the neutrino magnetic moment arising from a nonzero
neutrino mass\cite{fuji}, $\mu^{sm}_\nu=3.2\times10^{-19} m_\nu({\rm
eV})\mu_B$ (referred to as ``the extended standard electroweak theory'').
Accordingly, the upper limit of $\mu_{\nu}$ for the first 
generation neutrino is $\mu^{sm}_\nu\leq2.3\times10^{-18}\mu_B$ with
$m_\nu\leq7.3$ eV. The upper limit may also be obtained from leptonic
scatterings, which is typically $10^{-10}\mu_B$, or
from astrophysics studies with a more stringent upper limit of $10^{-11}\mu_B$.
Our numerical results for the first generation are summarized in Fig. 2, where 
the neutrino magnetic moment $\mu_\nu$ in units of $\mu_B$ is shown as a
function of the leptoquark mass. We note that, for the leptoquark mass 
$[V^{({2\over3})}_{1\mu}]$ of 50 to 100 TeV, $\mu_\nu$ is of order
$10^{-18}\mu_B$, a value compatible with the extended standard electroweak
theory.  

The upper limit of $\mu_{\nu}$ for the second generation neutrino is $0.51
\times 10^{-13}\mu_B$ (with $m_\nu\leq0.17$ MeV)
in the extended standard electroweak theory \cite{pdg}, 
or in the range of $10^{-10}\mu_B$ from leptonic scatterings,
while from astrophysics the typical value is $10^{-11}\mu_B$.                   
In Fig. 3, we describe our prediction on the neutrino magnetic moment
$\mu_\nu$ in units of $\mu_B$ as a function of the leptoquark mass of     
1 to 100 TeV. We obtain $\mu_\nu$ around $10^{-12}\sim 10^{-16}\mu_B$, a value
very close to being observable. 

The upper limit of $\mu_{\nu}$ for the third generation neutrino is $1.1 \times
10^{-11}\mu_B$ (with $m_\nu\leq35$ MeV) in the extended standard electroweak theory \cite{pdg}, or in
the range of $10^{-6}\sim10^{-7}\mu_B$ from leptonic scatterings, while from
astrophysics studies the upper limit is $10^{-12}\sim10^{-11}\mu_B$.         
In Fig. 4, we plot the third generation neutrino magnetic moment             
$\mu_\nu$ in units $\mu_B$ as a function of the leptoquark mass in the range of
1 to 100 TeV. We find that $\mu_\nu$ is of order $10^{-10}\sim 10^{-14}\mu_B$.

\section{Conclusion}

Vector leptoquarks in the TeV mass range, when couple to both left- and 
right-handed neutrinos, offer an alternative mechanism for generating a
nonvanishing neutrino magnetic moment, which in some cases is by no means
negligible. This alternative mechanism (which does not require a nonzero
neutrino mass) makes use of the special feature that leptoquarks couple
simultaneously to leptons and quarks. For the third generation neutrino, there
is a potential enhancement from the very large top quark mass making the
corresponding predicted neutrino magnetic moments fairly sizable.

\section{Acknowledgments}
We would like to acknowledge Dr. C.-T. Chan for valuable discussions. This work
was supported in part by a grant from National Science Council of Republic of
China (NSC88-2112-M002-001Y).
%
%

\begin{figure}[htb]                                                             
\centerline{\DESepsf(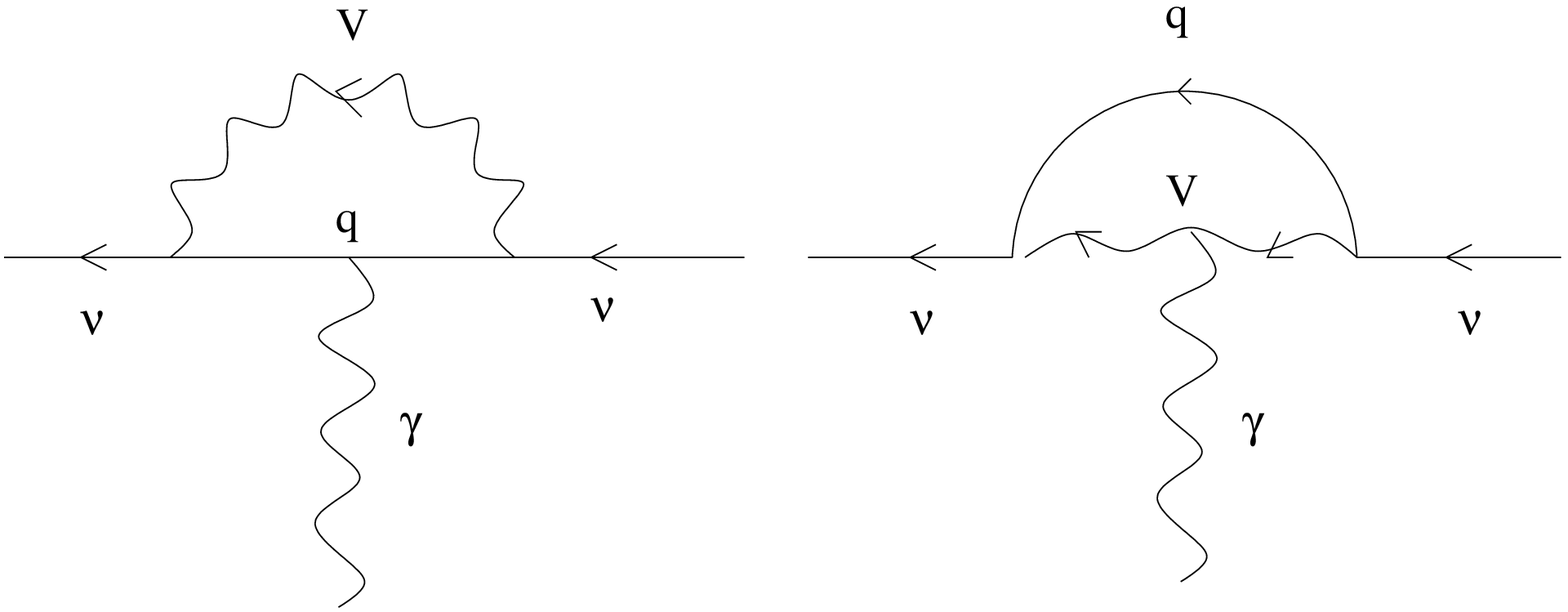 width 12cm)}                                        
\smallskip                                                                      
\caption{One loop diagrams which give rise to a nonzero neutrino magnetic
moment.}                  
\end{figure}  

\begin{figure}[htb]
\centerline{\DESepsf(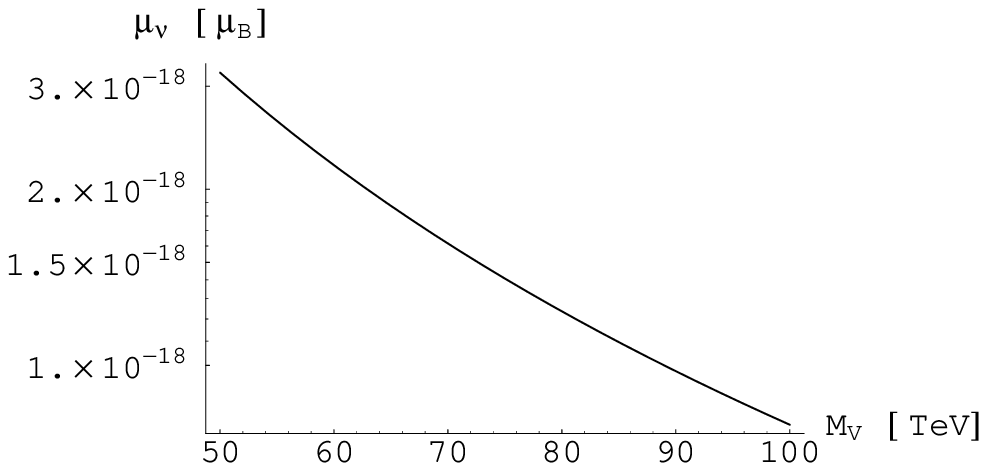 width 12cm)}
\smallskip
\caption{The first generation neutrino magnetic moment $\mu_\nu$ in units of
$\mu_B$ plotted as a function of the vector leptoquark mass in the range of 
50 to 100 TeV.} 
\end{figure}

\begin{figure}[htb]
\centerline{\DESepsf(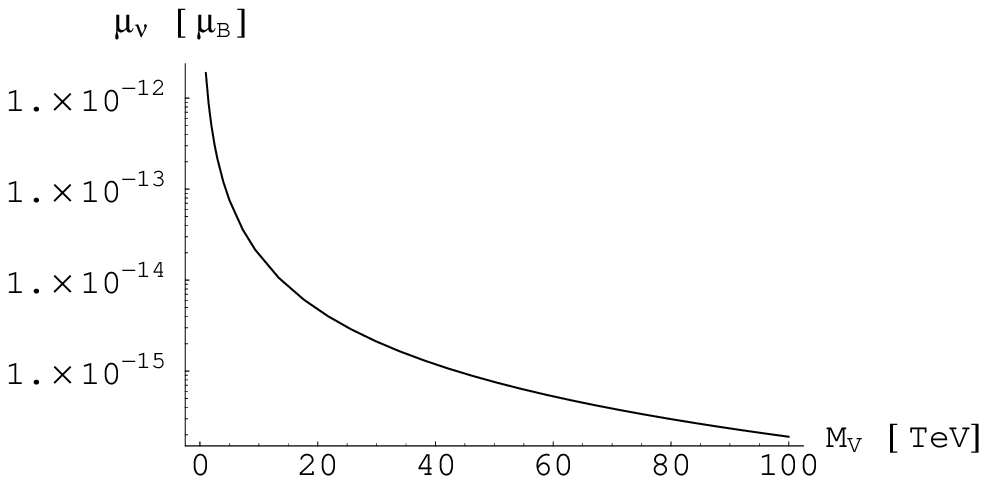 width 12cm)}
\smallskip
\caption{The second generation neutrino magnetic moment $\mu_\nu$ in units of
$\mu_B$ plotted as a function of the vector leptoquark mass in the range of 
1 to 100 TeV.}
\end{figure}

\begin{figure}[htb]
\centerline{\DESepsf(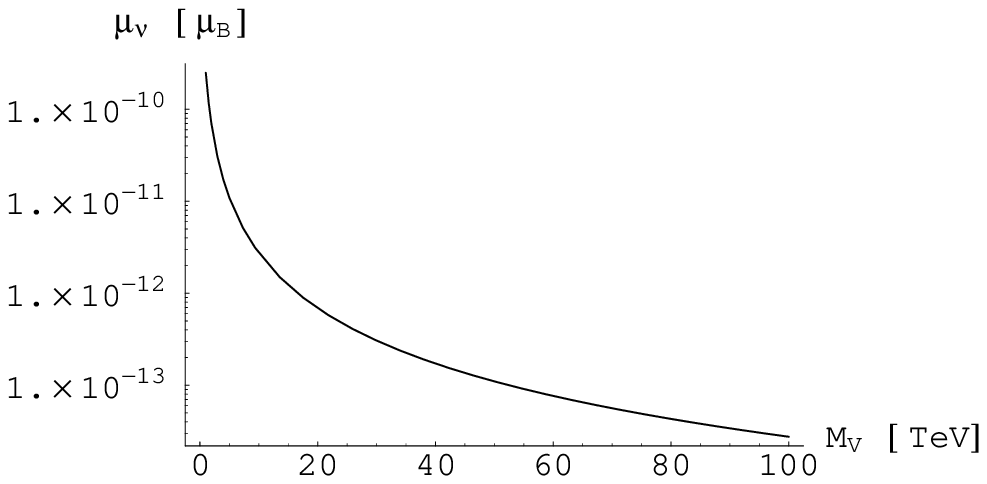 width 12cm)}
\smallskip
\caption{The third generation neutrino magnetic moment $\mu_\nu$ in units of
$\mu_B$ plotted as a function of the vector leptoquark mass in the range of 
1 to 100 TeV.}
\end{figure}
\end{document}